\documentclass[12pt]{article}
\usepackage{amsmath}
\usepackage{amssymb}
\begin{document}
\vskip 2cm
\begin{center}
{\sf {\Large  General $\mathcal{N }= 2$ Supersymmetric Quantum Mechanical Model: Supervariable 
Approach to its Off-Shell Nilpotent Symmetries}}

\vskip 3.5cm

{\sf S. Krishna$^{(a)}$, A. Shukla$^{(a)}$, R. P. Malik$^{(a,b)}$}\\
$^{(a)}$ {\it Physics Department, Centre of Advanced Studies,}\\
{\it Banaras Hindu University, Varanasi - 221 005, (U.P.), India}\\
$^{(b)}$ {\it DST Centre for Interdisciplinary Mathematical Sciences,}\\
{\it Faculty of Science, Banaras Hindu University, Varanasi - 221 005, India}\\
{\small {\sf {e-mails: skrishna.bhu@gmail.com; ashukla038@gmail.com;  rpmalik1995@gmail.com}}}

\end{center}

\vskip 3cm

\noindent
{\bf Abstract:} 
Using the supersymmetric (SUSY) invariant restrictions on the (anti-)chiral supervariables, we derive the off-shell nilpotent
symmetries of the {\it general} one (0 + 1)-dimensional $\mathcal{N }= 2$ SUSY quantum mechanical (QM) model which is considered on a 
(1, 2)-dimensional supermanifold (parametrized by a bosonic variable $t$ and a pair of Grassmannian variables $\theta$ and $\bar\theta$
 with $\theta^2 = \bar\theta^2 = 0$, $\theta\bar\theta + \bar\theta\theta = 0$). We provide the geometrical meanings to the
{\it two} SUSY transformations of our present theory which are valid for any arbitrary  
 type of superpotential. We express the conserved charges  and Lagrangian of  the theory in terms of the supervariables (that are obtained 
after the application of  SUSY invariant restrictions) and provide the geometrical interpretation for the nilpotency property  
and SUSY invariance of the Lagrangian for the general $\mathcal{N }= 2$ SUSY quantum theory. We also comment on the mathematical interpretation 
of the above symmetry transformations.\\



\newpage

\noindent
\section {Introduction}

The local non-Abelian 1-form
gauge theories are at the heart of theoretical description of three out of  four fundamental interactions of nature.
The Becchi-Rouet-Stora-Tyutin (BRST) formalism is one of the most intuitive approaches for the covariant canonical quantization 
of the $p$-form ($p = 1, 2, 3,...$) gauge theories where the local gauge symmetry of the original 
theory is traded with the ``quantum" gauge 
[i.e. (anti-)BRST] symmetries. A couple of decisive features of the (anti-)BRST symmetries are their nilpotency and absolute
anticommutativity.  These mathematical properties are explained 
{\it geometrically} by the well-known superfield formalism [1-8]
where the horizontality condition (HC) plays a key role.
The HC is important {\it only} in the context of BRST description of the $p$-form (non-)Abelian theories
where there is {\it no}  coupling between  the gauge and  matter fields.

For the interacting gauge theories, one requires more restrictions than the celebrated HC. In a set of papers (see, e.g. [9-12]),
the additional gauge invariant restrictions (GIRs) have been exploited, besides HC, to obtain the {\it full} set of off-shell nilpotent and absolutely
anticommuting (anti-)BRST symmetries for the gauge, matter and (anti-)ghost fields of a given interacting gauge theory. 
It has been a {\it challenging} problem to apply the appropriate form of the above superfield formalisms [1-12] 
to the supersymmetric (SUSY) theories where the nilpotency property exists but the absolute 
anticommutativity  property {\it does} not. In our present paper, we address this problem in 
the context of $\mathcal{N} = 2$ SUSY quantum mechanical (QM) model which happens to be a one 
(0 + 1)-dimensional (1D) SUSY system. We observe that it is the generalization of the idea of GIRs [9-12]
to the SUSY invariant restrictions that plays a key role in our whole theoretical discussions.

The central theme of our present investigation is to exploit the strength of SUSY invariant restrictions on the (anti-)chiral supervariables to capture
the nilpotency property of the SUSY symmetry transformations for the general $\mathcal{N} = 2$ SUSY QM model and derive the full set of SUSY
symmetries in an accurate manner. We also provide the {\it geometrical} basis for the SUSY symmetry invariance of the Lagrangian of the $\mathcal{N} = 2$
SUSY QM system. We lay emphasis on the fact that, to {\it avoid} the absolute anticommutativity property of the $\mathcal{N} = 2$ SUSY transformations, we are theoretically compelled to choose 
the (anti-) chiral supervariables defined on the (1, 1)-dimensional super-submanifolds of the full (1, 2)-dimensional supermanifold. 
The latter  is parameterized by the superspace coordinate $Z^M = (t, \theta, \bar\theta)$ with a pair of Grassmannian variables $\theta$
and $\bar\theta$ (with $\theta^2 = 0, \;\bar\theta^2 = 0, \;\theta \,\bar\theta + \bar\theta \,\theta = 0 $)
and an evolution parameter $t$.

The (anti-)BRST and  $\mathcal{N} = 2$ SUSY symmetry transformations are nilpotent 
of order two. However, they differ drastically in 
their anticommutativity property. Whereas the former symmetries turn out to 
be absolutely anticommuting, the latter symmetries 
do {\it not} obey the same rule. The similarity  between the above two types  of symmetries is only at the level of nilpotency. The latter   could be of off-shell
and/or on-shell variety. Furthermore, the $\mathcal {N} = 2$ SUSY transformations are {\it two} 
in number as is the case with nilpotent
 BRST  and anti-BRST symmetries for a given local gauge transformation of a (non-)Abelian
$p$-form gauge theory. We discuss these issues (i.e. nilpotency and anticommutativity properties) 
in our Appendix A, too.

Within the framework  of superfield approach to BRST formalism [1-12], the 
nilpotent symmetries have been identified with the
translational generators ($\partial_\theta, \, \partial_{\bar\theta}$) 
along the Grassmannian directions of the (D, 2)-dimensional supermanifold on which
a given D-dimensional (non-)Abelian gauge theory is generalized. In this approach, 
the superfields are expanded along {\it all}
($1, \theta, \bar\theta, \theta\bar\theta$) directions of the (D, 2)-dimensional supermanifold.  
The key features of
this expansion are the fact that we obtain the off-shell nilpotent and absolutely anticommuting
(anti-)BRST symmetries for a given D-dimensional gauge theory due to HC and GIRs
on the superfields. The nilpotency and absolute anticommutativity owe their origin to the properties
$\partial^2_\theta = \partial^2_{\bar\theta} = 0$ and $\partial_\theta\,\partial_{\bar\theta} + \partial_{\bar\theta}\,\partial_{\theta} = 0$,
respectively, of the translational generators $\partial_{\bar\theta}$ and $\partial_\theta$.
The (anti-)chiral 
supervariables have been chosen in   our present theory so that the translational 
generators ($\partial_{\bar\theta}$ or $\partial_\theta$) along the Grassmannian 
direction ($\bar\theta$ or $\theta$) of the (1, 1)-dimensional (anti-)chiral supermanifolds 
could be identified with {\it one} of the {\it two} $\mathcal{N} = 2$ SUSY transformations. 
This choice, in fact, enables us to capture {\it only} the nilpotency property of   
the $\mathcal{N} = 2$ SUSY transformations
and it {\it avoids} any assertion about the anticommutativity property (see also Appendix A).

The contents of our present endeavor are organized as follows. In Sec. 2, we discuss the continuous symmetries of the Lagrangian
for the general $\mathcal{N} = 2$ SUSY QM model.  Our Sec. 3 contains the discussion on the derivation of the {\it first} SUSY symmetry by imposing the appropriate 
SUSY invariant  restrictions (SUSYIRs) on the anti-chiral supervariables. Our Sec. 4 is devoted to the derivation of the {\it second} SUSY 
symmetry  from the SUSYIRs on the chiral supervariables. In the forthcoming  Sec. 5, we deal with the proof of nilpotency of the SUSY
transformations and invariance of the Lagrangian in the language of supervariables. 
We provide mathematical interpretation of the off-shell nilpotent 
$\mathcal{N} = 2$ SUSY transformations in the language of cohomological operators
of differential geometry in Sec. 6. Finally, we make some concluding remarks  in our Sec. 7.

 In our Appendix A, we provide convincing and cogent reasons behind our choice  of 
the (anti-)chiral supervariables.

\section {Preliminaries: General $\mathcal{N }= 2$ SUSY QM System with Any Arbitrary Superpotential}

Let us begin with the Lagrangian for the {\it general}  $\mathcal{N} = 2$ SUSY 
QM model as follows (see, e.g. [14])  
\begin{eqnarray}
L_0 = \frac{1}{2}\,\dot x^2 + i\, \bar\psi\,\dot\psi + W' A + \frac{1}{2}\, A^2 + W''\, \bar\psi\, \psi ,
\end{eqnarray}
where the overdot  and primes (i.e. $\dot x = dx/dt, \,\dot\psi = d\psi/dt$, 
\,$W^{\prime}(x) = dW/dx,\, W^{\prime\prime} = d^2 W/dx^2$) are
the notations for the time derivative and space derivatives, respectively. Here $x(t)$ is the bosonic variable and its
$\mathcal{N} = 2$ SUSY fermionic ($\psi^2 = {\bar\psi}^2= 0,\, \psi\,\bar\psi + \bar\psi\,\psi = 0$) counterparts are $\psi(t)$ and
$\bar\psi(t)$.  The evolution parameter in our theory is
$t$  and {\it classically} we have the absolute anticommutativity property 
between the fermionic variables $\psi$ and $\bar\psi$. 
The superpotential $W(x)$ is usually an even function of $x(t)$ [i.e. $W(-x) = W(x)$] and is {\it not} explicitly dependent on the evolution parameter $t$. 
This function is arbitrary for the case of {\it general} $\mathcal{N} = 2$ SUSY QM model and the auxiliary variable $A(t)$ is connected 
[i.e. $A(t) = - W^{\prime}(x)$] with the space derivative  on the superpotential $W(x)$. The above
Lagrangian is actually derived from the general $\mathcal{N} = 2$ superspace approach (see, e.g. [15,16] for details)
to SUSY quantum mechanics and its form is quite general.

Our theory being $\mathcal{N} = 2$ SUSY QM model, we have the following {\it two} nilpotent ($s^2_1 = s^2_2 = 0$)
 SUSY transformations $s_1$ and $s_2$ (see, e.g. [14]): 
  \begin{eqnarray}
&& s_1 x = i\, \psi, \qquad s_1 \psi = 0, \qquad s_1 \bar\psi = -(\dot x + i\, A),\qquad s_1\, A = -\dot \psi, \nonumber\\&&  s_2 x = i\,\bar\psi, \qquad s_2 \bar\psi = 0, \qquad s_2 \psi = -  (\dot x - i\, A),
\qquad s_2\, A =  \dot {\bar\psi},
\end{eqnarray}
under which the Lagrangian (1) transforms to the total time derivatives, as:
\begin{eqnarray}
s_1\, L_0 = \frac {d}{dt}\, \Big[- W'\, \psi \Big],  \quad\qquad
 s_2\, L_0 = \frac {d}{dt} \,\Big[i\, \bar\psi (\dot x - i A) + \bar\psi \,W'\Big].
\end{eqnarray}
As a consequence, the action integral ($S = \int dt\, L_0$) remains invariant.  It should be noted that $s_1$ and $s_2$ are off-shell 
nilpotent ($s^2_1 = s^2_2 = 0$) because we do {\it not} use {\it anywhere} the following Euler-Lagrange (EL) equations of motion:
\begin{eqnarray}
&&\ddot{\psi} + (W'')^2\, \psi - i\, W'''\, \dot {x} \, \psi = 0,
\;\;\quad \ddot x = W''\, A + W'''\, \bar\psi\,\psi, 
 \;\;\quad \dot{\psi} = i\, W''\, \psi,\nonumber\\
&& \ddot{\bar\psi}  + (W'')^2\, \bar\psi + i\, W'''\, \dot {x} \, \bar\psi = 0,
\qquad  A = - W', \qquad \dot{\bar\psi} = -i\, W''\, \bar\psi,
\end{eqnarray}
(emerging from the Lagrangian (1)) in the proof of their nilpotency.

According to Noether's theorem, the invariance of the action integral  leads to the derivation of the conserved charges as listed below:
\begin{eqnarray}
 Q = (i \,\dot x - A)\, \psi \equiv (i\, p - A)\, \psi, \qquad\quad  \bar Q = \bar\psi\,(i \,\dot x + A) \equiv \bar\psi \,(i\, p + A),
\end{eqnarray}
where $p = \dot x$ is the momentum corresponding to the bosonic variable. These charges turn out to
be the generators of transformations $s_1$ and $s_2$ because we have the following: 
\begin{eqnarray}
&& s_1\, \Phi = -i\, [\Phi,\; Q]_{\pm}, \qquad  s_2\, \Phi = -i\, [\Phi,\; \bar Q]_{\pm}, \qquad
 \Phi = x, \, \psi,\, \bar\psi,
\end{eqnarray}
where the subscripts ($\pm$), on the square bracket, correspond to the (anti)co-mmutator for the generic variable $\Phi$ being
(fermionic)bosonic in nature. The above charges $Q$ and $\bar Q$ are conserved ($\dot Q = \dot{\bar Q} = 0$) 
as can be directly checked by using the EL equations of motion (4).

One of the decisive features of the general $\mathcal{N} = 2$ SUSY QM model is the observation 
that the anticommutator of $s_1$ and $s_2$
should {\it not} be  zero and it must generate the time translation. 
This can be checked to be true in our theory as we have the following:
\begin{eqnarray}
&& \{s_1,\, s_2\}\, \Phi = s_\omega\, \Phi = (-\,2i)\,\dot\Phi, 
 \qquad \quad s_\omega = \{s_1, \, s_2\},
\nonumber\\ && \Phi = x,\, \psi,\, \bar\psi,\, A, \,W',\, W''.
\end{eqnarray}
The above equation establishes that the two successive operations of 
SUSY transformations $s_1$ and $s_2$ leads to the time derivative on a 
specific variable of the theory [modulo a factor of ($-\,2i$)]. 
Thus, we have the continuous symmetry transformation $s_\omega$ 
that transforms $L_0$  as:
\begin{eqnarray}
s_\omega\, L_0 = (s_1\,s_2 + s_2\,s_1)\, L_0 = \frac{dL_0}{dt}.
\end{eqnarray}
According to Noether's theorem, this continuous transformation,
 too, leads to the derivation of a conserved charge $Q_\omega$ as:
\begin{eqnarray}
Q_\omega = \frac{p^2}{2} - \frac{1}{2}\, A^2 - A\, W' - W'' \bar\psi\, \psi \equiv  H,
\end{eqnarray}
where $H$ is the Hamiltonian of the theory.

\section{Off-Shell Nilpotent Continuous  SUSY Symmetry Transformations: Anti-chiral Supervariables}

It is clear from (8) that the $\mathcal{N} = 2$ SUSY transformations $s_1$ and $s_2$ are {\it not}
absolutely anticommuting. Thus, to derive the SUSY transformations $s_1$, we have to concentrate 
on the (1, 1)-dimensional super-submanifold 
that is parametrized by  the superspace variables ($t, \bar\theta$). We have to impose SUSY
invariant restrictions on the anti-chiral supervariables which are function of ($t, \bar\theta$)
{\it only}. The first step, towards our main goal of deriving $s_1$, is to generalize all the ordinary
(explicitly time-dependent) variables [cf. (1)] to their counterpart supervariables as 
\begin{eqnarray}
&&x(t) \; \longrightarrow \; X(t, \theta, \bar\theta)\mid_{\theta = 0}\,
 \equiv X(t,\bar\theta) = x(t) +  \bar\theta\, f_1(t),\nonumber\\
&&\psi(t) \;\longrightarrow \; \Psi (t, \theta, \bar\theta) \mid_{\theta = 0}\, 
\equiv \Psi (t,  \bar\theta) = \psi (t) 
 + i\, \bar\theta\, b_1 (t), \nonumber\\
&&\bar\psi(t) \;\longrightarrow \; \bar\Psi (t, \theta, \bar\theta)\mid_{\theta = 0}\, 
\equiv \bar\Psi (t,  \bar\theta) = \bar\psi (t) 
 + i\, \bar\theta\, b_2 (t),\nonumber\\
&& A(t)  \longrightarrow {\tilde A}(t, \theta, \bar\theta)\mid_{\theta = 0}\,
 \equiv {\tilde A}(t, \bar\theta) = A(t) + \bar\theta\, f_2(t),
\end{eqnarray} 
where the pair of secondary variables ($b_1, b_2$) and ($f_1, f_2$) are bosonic and 
fermionic in nature, respectively. We  also observe that
the total number of bosonic ($x, A, b_1$, $b_2$) and fermionic ($\psi, \bar\psi, f_1, f_2$) 
variables (and their corresponding degrees of freedom)
do match which is one of the basic requirements of any arbitrary general SUSY theory.
The expansion (10) should be contrasted with the 
expansions that are used in the context of BRST formalism where the superfields are expanded
along {\it all} the Grassmannian directions ($1, \theta, \bar\theta, \theta\bar\theta$) 
of the (D, 2)-dimensional supermanifold (cf. Appendix A) for a given
D-dimensional gauge theory [9-12].

It is obvious from (2) that   $s_1 \psi = 0$. Hence, the fermionic variable $\psi$ is a 
SUSY invariant quantity under $s_1$. 
We demand that this quantity should remain independent of the ``soul" variable $\bar\theta$. 
As a consequence, we have the SUSY invariant restriction  
\begin{eqnarray}
\Psi (t, \theta, \bar\theta) \mid_{\theta = 0}\, \equiv \,\Psi (t, \bar\theta) = \psi (t) \quad\Longrightarrow \quad b_1(t) = 0.
\end{eqnarray}
Furthermore, we note that  $s_1\,(x\,\psi) = 0$ and $s_1(\dot {x} \,\psi) = 0$ (primarily   due to the fermionic nature of $\psi$ where $\psi^2 = 0$). 
Thus, we {\it also} have the other SUSY restrictions as 
\begin{eqnarray}
&& X(t, \bar\theta)\,\Psi (t, \bar\theta) = x(t)\, \psi (t),\qquad\quad
 \dot{X}(t, \bar\theta)\,\Psi (t, \bar\theta) = \dot{x}(t)\, \psi (t).
\end{eqnarray}
Using the result from (11), we obtain (from the above SUSY restrictions) the following:
\begin{eqnarray}
f_1(t)\,\psi (t) = 0,\qquad\qquad \dot f_1(t)\,\psi (t) = 0.
\end{eqnarray}
The non-trivial solution of the above restrictions is  $f_1(t)$ $\propto \psi (t)$.
For the algebraic convenience, however,  
we choose $f_1(t) = i\,\psi(t)$ which will be useful later [cf. (19) below].

It is worthwhile to mention here about the analogy between the above restrictions 
and the gauge invariant restrictions exploited in the context of ``augmented" superfield approach
to BRST formalism [see, e.g. (9-12)]. In fact, in the latter approach, the gauge invariant (physical) 
quantities are taken to be independent of the
``soul" (i.e. Grassmannian) variables because the latter are merely a mathematical 
artifact and they have {\it no} physical realizations. This requirement leads,
in particular, to the precise derivation of the (anti-)BRST symmetries for the matter fields
in an interacting gauge theory (see, e.g. [9-12] for details).
This idea has been promoted in our SUSY invariant theory where we have tapped 
the potential and power of SUSY invariant restrictions.

A close look at the transformations (2) shows that the nilpotency of 
$s_1$ [i.e. $s^2_1\, \bar\psi = -\, s_1 (\dot x + i\,A)= 0$]
implies that we  have a SUSY invariant quantity ($\dot x + i\, A$) under $s_1$. 
Thus, we impose the following SUSY invariant restriction
\begin{eqnarray}
\dot {X}(t, \bar\theta) + i\, \tilde A(t, \bar\theta) = \dot {x}(t) + i \,A (t),
\end{eqnarray} 
which leads to the relationship $f_2 + \dot {\psi} = 0$. This implies that $f_2 = -\,\dot{\psi}$.
Finally, from the symmetry invariance of $L_0$, we observe that the following specific combination
\begin{eqnarray}
\frac{1}{2} \,\dot x^2(t)+ i\, \bar\psi (t)\, \dot\psi (t) + \frac{1}{2}\, A^2(t) \equiv C(t),
\end{eqnarray}
is a SUSY invariant quantity (i.e. $s_1 C(t) = 0 $). Thus, we have the following SUSY invariant restriction that 
incorporates sum of the composite (super)variables:
\begin{eqnarray}
&&\frac{1}{2}\,\dot {X}^2(t, \bar\theta) + i\, \bar\Psi (t, \bar\theta)\, \dot{\Psi}(t, \bar\theta) 
+  \frac{1}{2}\, {\tilde A}^2(t, \bar\theta) \nonumber\\ &&
= \frac{1}{2} \,\dot x^2(t)+ i\, \bar\psi (t)\, \dot\psi (t) + \frac{1}{2}\, A^2(t).
\end{eqnarray}
The above restriction leads to the following relationship
\begin{eqnarray}  
\dot x \,{\dot f}_1 - b_2 \,\dot\psi + f_2 \,A = 0.
\end{eqnarray}
The substitution of $f_1 = i\,\psi$ and $f_2 = - \dot \psi$ in the above, implies the following
\begin{eqnarray}
b_2 = i\, \dot x - A.
\end{eqnarray}
We conclude that the SUSY restrictions (11), (12), (14) and (16) lead to the following
expansions of the anti-chiral supervariables [cf. (10)] as:
\begin{eqnarray}
&&X^{(1)}(t,\theta, \bar\theta)\mid_{\theta = 0} \,=  X^{(1)}(t, \bar\theta),\nonumber\\ &&
X^{(1)}(t, \bar\theta) = x(t) +  \bar\theta\, (i\,\psi) \equiv x(t) + \bar\theta \,(s_1\, x),\nonumber\\
&&\Psi^{(1)} (t, \theta,\bar\theta) \mid_{\theta = 0}\, = \Psi^{(1)} (t, \bar\theta), \nonumber\\ &&
\Psi^{(1)} (t, \bar\theta) = \psi (t)  + \bar\theta\,(0) \equiv \psi(t) + \bar\theta\, (s_1\, \psi), \nonumber\\
&&\bar\Psi^{(1)} (t,\theta, \bar\theta)\mid_{\theta = 0}\, = \bar\Psi^{(1)} (t, \bar\theta), \nonumber\\ &&
 \bar\Psi^{(1)} (t, \bar\theta) = \bar\psi (t)  + \, \bar\theta\, (-\dot x - i A) \equiv \bar\psi (t)  
+ \, \bar\theta\, (s_1\, \bar\psi),\nonumber\\
&&{\tilde A}^{(1)}(t, \theta, \bar\theta) \mid_{\theta = 0}\, = {\tilde A}^{(1)}(t, \bar\theta), \nonumber\\ &&
{\tilde A}^{(1)}(t, \bar\theta) = A(t) +  \bar\theta\, (-\dot\psi) \equiv A(t) + \bar\theta \,(s_1\, A).
\end{eqnarray} 
Here the superscript $(1)$ denotes the expansions of supervariables obtained after the application of
SUSY invariant restrictions. Thus, we have derived the SUSY transformations $s_1$ [cf. (2)] 
in a very clear fashion using the SUSY invariant restrictions on the anti-chiral supervariables.

From the expansion (19), it is clear that we have the following relationship
between the Grassmannian derivative $\partial_{\bar\theta}$ and the SUSY transformations $s_1$, namely; 
\begin{eqnarray}
\frac{\partial}{\partial\bar\theta}\, \Omega^{(1)} (t, \theta, \bar\theta) \mid_{\theta = 0}\, \equiv 
\frac{\partial}{\partial\bar\theta}\, \Omega^{(1)} (t, \bar\theta) = s_1\, \Omega (t),
\end{eqnarray}
where $\Omega^{(1)} (t, \bar\theta)$ is the generic supervariable obtained after the application of 
SUSY restriction and $\Omega (t)$ is generic variable in the one (0 + 1)-dimensional ordinary space.
Geometrically, it is clear that the SUSY transformations ($s_1$) for a generic one (0 + 1)-dimensional 
variable $\Omega (t)$ is equivalent to the translation of its corresponding supervariable $\Omega^{(1)} (t, \bar\theta)$ 
 along the
$\bar\theta$-direction of super-submanifold where the anti-chiral supervariables are defined.
In view of the definition of the generator (i.e. $s_1\,\Phi = -i\,[\Phi,\, Q]_{\pm}$), it is obvious
that the translational generator $\partial_{\bar\theta},$ along the $\bar\theta$-direction 
of the (1, 1)-dimensional anti-chiral super-submanifold, is {\it also} connected with the super charge
$Q$. Finally, we have the mapping   
$\partial_{\bar\theta} \leftrightarrow s_1 \leftrightarrow Q $ where the nilpotency  property of the
operators ($s_1,\, Q,\, \partial_{\bar\theta}$) is intertwined in a beautiful fashion as they are 
inter-dependent on one-another.

We wrap up this section with the remark that the generic supervariable $ \Omega^{(1)} (t,  \bar\theta)$ 
is actually the anti-chiral limit of the most general supervariable $\Omega^{(1)} (t, \theta, \bar\theta)$ 
[i.e. $\Omega^{(1)} (t, \theta, \bar\theta)\mid_{\theta = 0}\,
 \equiv \Omega^{(1)} (t,  \bar\theta)$ defined on the general (1, 2)-dimensional supermanifold].
This is precisely the reason that  we have continued with the 
partial nature of the Grassmannian derivative $\partial_{\bar\theta}$ and have {\it not} taken
 the total derivative (i.e. $d/d\bar\theta$) w.r.t. $\bar\theta$ in the relationship (20).

\section{Off-Shell Nilpotent Continuous SUSY Symmetry Transformations: Chiral Supervariables}

To derive the SUSY transformations $s_2$, we focus on the chiral super-submanifold 
which is parametrized by the superspace variables ($t, \theta$). 
The basic and auxiliary variables (depending explicitly on $t$) of the Lagrangian (1) are, first 
of all, generalized onto the (1, 1)-dimensional chiral super-submanifold as:   
\begin{eqnarray}
&&x(t) \; \longrightarrow \; X(t, \theta, \bar\theta)\mid_{\bar\theta =0} \,\equiv X(t, \theta)= x(t)
 +  \theta\, {\bar f}_1(t),\nonumber\\
&&\psi(t) \;\longrightarrow \; \Psi (t, \theta, \bar\theta)\mid_{\bar\theta =0}\,
\equiv \Psi (t, \theta)= \psi (t)  + i\,\theta\,{\bar b}_1 (t), \nonumber\\
&&\bar\psi(t) \;\longrightarrow \; \bar\Psi (t, \theta, \bar\theta)\mid_{\bar\theta =0}\,
\equiv \bar\Psi (t, \theta) = \bar\psi (t)  + i\, \theta\, {\bar b}_2 (t),\nonumber\\
&& A(t)  \longrightarrow  {\tilde A}(t, \theta, \bar\theta)\mid_{\bar\theta =0}
\equiv {\tilde A}(t, \theta) = A(t) + \theta\, {\bar f}_2(t),
\end{eqnarray} 
where the secondary variables (${\bar f}_1,\, {\bar f}_2$) are fermionic and their  counterparts 
(${\bar b}_1,\, {\bar b}_2$) are bosonic in nature. On the r.h.s. of (21), we observe that the fermionic
(${\bar f}_1,\, {\bar f}_2,\, \psi,\, \bar\psi$) degrees of freedom match with their counterpart bosonic 
(${\bar b}_1,\, {\bar b}_2,\, x,\, A$) degrees of freedom.

The above secondary variables (${\bar b}_1,\, {\bar b}_2,\, {\bar f}_1,\, {\bar f}_2$) can be determined in terms of 
the basic variables if we impose the proper SUSY invariant restrictions on the chiral supervariables. For instance,
we observe that $s_2\, \bar\psi = 0$ [cf. (2)]. Thus, we impose the SUSY restriction 
\begin{eqnarray}
\bar\Psi (t,\theta, \bar\theta)\mid_{\bar\theta = 0} \,\equiv \,\bar \Psi (t, \theta) = \bar \psi (t) 
\quad\Longrightarrow \quad {\bar b}_2(t) = 0.
\end{eqnarray}
We also note that $s_2\, (x\, \bar\psi) = 0$, $s_2\, (\dot x\, \bar\psi) = 0$ [cf. (2)]
because of the fermionic nature of $\bar\psi$.
As a result, we have the following two SUSY restrictions on the composite chiral supervariables:
\begin{eqnarray}
&& X(t, \theta)\,\bar \Psi (t, \theta) = x(t)\, \bar\psi (t), \qquad\quad
{\dot X}(t, \theta)\,\bar \Psi (t, \theta) = {\dot x}(t)\, \bar\psi (t).
\end{eqnarray}
With the help from (22), we find that 
\begin{eqnarray}
{\bar f}_1(t)\,\bar\psi (t) = 0, \qquad\qquad \dot {\bar f}_1(t)\,\bar\psi (t) = 0.
\end{eqnarray}
The non-trivial solution of the above restrictions is ${\bar f}_1 = i\, \bar\psi$. We have taken $i$ factor 
for the algebraic convenience which will become clear later [cf. (28) below] in our further discussions.

To determine all the secondary variables, we note further that $s_2\, [\dot {x}(t) - i A (t)] = 0$.
This invariance emerges from the nilpotency of $s_2$ because we observe that 
$s^2_2\, \psi = s_2\,(-\,[\dot {x} - i A ]) = 0 $
in equation (2). This shows that ($\dot x - i\, A$) is a SUSY invariant quantity. Thus, we have the following SUSY invariant 
restriction on the chiral supervariables: 
\begin{eqnarray}
\dot {X}(t, \theta) - i \,\tilde A(t, \theta) = \dot {x}(t) - i\, A (t).
\end{eqnarray}
The above condition yields ${\bar f}_2 = \dot{\bar\psi}$. A part of the {\it modified} form of Lagrangian (1)
also remains invariant under $s_2$. In fact, we note that the following sum of the composite
variables are invariant under $s_2$, namely; 
\begin{eqnarray}
s_2\, \Big[\frac{1}{2} \,\dot x^2(t)-  i\, \dot{\bar\psi} (t)\, \psi (t) + \frac{1}{2}\, A^2(t)\Big] = 0.
\end{eqnarray}
Thus, we have the following SUSY invariant restriction on the specific combination of 
composite chiral supervariables:
\begin{eqnarray}
&& \frac{1}{2}\,\dot {X}^2(t, \theta) - i\, \dot{\bar\Psi} (t, \theta)\, {\Psi}(t, \theta) 
+  \frac{1}{2}\, {\tilde A}^2(t, \theta)\nonumber\\ &&
= \frac{1}{2} \,\dot x^2(t)- i\, \dot{\bar\psi} (t)\, \psi (t) + \frac{1}{2}\, A^2(t),
\end{eqnarray}
which leads to the determination of $\bar b_1 = i \dot x + A$.

Plugging in the value  ${\bar f}_1 = i \bar\psi,\, {\bar b}_2 = 0,\, {\bar f}_2 = \dot{\bar\psi}$ and  
$\bar b_1 = i \dot x + A$, we obtain the following expansions 
\begin{eqnarray}
&&X^{(2)}(t, \theta, \bar\theta)\mid_{\bar\theta =0} \,= X^{(2)}(t, \theta), \nonumber\\ &&
X^{(2)}(t, \theta) = x(t) +  \theta\, (i\,\bar\psi) \equiv x(t) + \theta \,(s_2\, x),\nonumber\\
&&\Psi^{(2)} (t, \theta, \bar\theta)\mid_{\bar\theta =0}\, = \Psi^{(2)} (t, \theta), \nonumber\\ &&
\Psi^{(2)} (t, \theta) = \psi (t)  + \theta\,(- \dot x + i A) \equiv \psi(t) + \theta\, (s_2\, \psi), \nonumber\\
&&\bar\Psi^{(2)} (t, \theta, \bar\theta)\mid_{\bar\theta =0}\, = \bar\Psi^{(2)} (t, \theta), \nonumber\\ &&
\bar\Psi^{(2)} (t, \theta) = \bar\psi (t)  + \theta\, (0) \equiv \bar\psi (t)  +  \theta\, (s_2\, \bar\psi),\nonumber\\
&&{\tilde A}^{(2)}(t, \theta, \bar\theta)\mid_{\bar\theta =0} \,=  \tilde A^{(2)}(t, \theta), \nonumber\\ &&
\tilde A^{(2)}(t, \theta) = A(t) +  \theta\, (\dot{\bar\psi}) \equiv A(t) + \theta \,(s_2\, A).
\end{eqnarray} 
Furthermore, we have found that the following  relationship is true, namely;
\begin{eqnarray}
\frac{\partial}{\partial\theta}\, \Omega^{(2)}(t, \theta, \bar\theta) \mid_{\bar\theta = 0} \,\equiv 
\frac{\partial}{\partial\theta}\, \Omega^{(2)}(t, \theta) = s_2\, \Omega (t).
\end{eqnarray}
The above relation demonstrates  that the translation of the generic chiral supervariable 
$\Omega^{(2)}(t, \theta) \equiv X^{(2)}(t, \theta)$, $
\Psi^{(2)} (t, \theta),\, \bar\Psi^{(2)} (t, \theta),\, \tilde A^{(2)}(t, \theta)$
along the Grassmannian direction $\theta$ of the chiral (1, 1)-dimensional super-submanifold generates
the SUSY transformations $s_2$ on the 1D ordinary generic variable $\Omega(t)$ [cf. (1)].
However, as we know from (6), $\bar Q$ is also the generator for $s_2$ because 
$s_2\, \Omega = -i\, [\Omega,\, \bar Q]_{\pm}$. Thus, we conclude that the following mapping
\begin{eqnarray}
\frac{\partial}{\partial \theta} \;\;\longleftrightarrow \;\; s_2 \;\;\longleftrightarrow \; \; \bar Q,
\end{eqnarray}
exists amongst the translation generator ($\partial_\theta$),  symmetry transformation ($s_2$) and  conserved charge $\bar Q$.
The nilpotency of $s_2$ (i.e. $s^2_2 = 0$) is also encoded in the nilpotency of SUSY
charge $\bar Q$ which, in turn, is deeply related to the nilpotency ($\partial^2_\theta = 0$)
of the Grassmannian derivative ($\partial_\theta $). Thus, the nilpotency of ($s_2, \bar Q, \partial_\theta$) are inter-related.
Within the framework of supervariable approach, the nilpotency  of $s_2$ and $\bar Q$ is encoded in the two successive
translations along $\theta$-direction [cf. (29), (30)] because $\partial_\theta^2 = 0$.

\section{Symmetry Invariance and Off-Shell Nilpotency: Supervariable Approach}

In this  section, we capture the symmetry invariance of the Lagrangian under SUSY transformations $s_1$ and $s_2$ 
and the off-shell nilpotency of the charges $Q$ and $\bar Q$ in the language of supervariables obtained 
after the application of SUSYIRs. Using the expansion (19), it can be seen that 
the Lagrangian (1) can be generalized [onto (1, 1)-dimensional chiral super-submanifold] in terms of the anti-chiral supervariables as: 
\begin{eqnarray}
 L_0 \; \Longrightarrow    \; {\tilde L}^{(ac)}_0 &=& \frac{1}{2}\,\dot {X^{(1)}}(t, \bar\theta)\,\dot {X^{(1)}}(t, \bar\theta) + i\, {\bar\Psi}^{(1)} (t, \bar\theta)\,{\dot{\Psi}}^{(1)}(t, \bar\theta) \nonumber\\  
&+& \frac{1}{2}\,{\tilde A}^{(1)}(t, \bar\theta)\,{\tilde A}^{(1)}(t, \bar\theta)
+ {\tilde W}^{\prime}(X^{(1)})\;{\tilde A}^{(1)}(t, \bar\theta) \nonumber\\  
&+& {\tilde W}^{\prime\prime}(X^{(1)}) \, {\bar\Psi}^{(1)} (t, \bar\theta)\,{{\Psi}}^{(1)}(t, \bar\theta),
\end{eqnarray}
where the superscript ($ac$) denotes the expression for the Lagrangian in terms of the anti-chiral supervariables.
It can be checked explicitly that:
\begin{eqnarray}  
&& {\tilde W}^{\prime}(X^{(1)}) = W^{\prime}(x) + \bar\theta \,\Big[i\, W^{\prime\prime}(x)\,\psi(t)\Big],
\nonumber\\ && {\tilde W}^{\prime\prime}(X^{(1)}) = W^{\prime\prime}(x) + \bar\theta \,\Big[i\, W^{\prime\prime\prime}(x)\,\psi(t)\Big],
\end{eqnarray}
where we have used the Taylor expansion with $X^{(1)}(t,\bar\theta) = x(t) + i\, \bar\theta\, \psi(t) \equiv x(t) + s_1 x(t)$ around $x(t)$.
In view of the mapping $s_1\leftrightarrow \partial_{\bar\theta}$ [cf. (20)],  we note that the invariance 
of the Lagrangian  (1) under $s_1$ can be expressed in the following fashion as the Grassmannian derivative on
${\tilde L}^{(ac)}_0$:
\begin{eqnarray}
\frac{\partial}{\partial \bar\theta}\, {\tilde L}^{(ac)}_0 = \frac{d}{dt} \,\bigl[- W^{\prime} \,\psi\bigr]
 \quad\Longleftrightarrow \quad  s_1\, L_0 = \frac{d}{dt} \,\bigl[- W^{\prime} \,\psi\bigr].
\end{eqnarray}
Geometrically, the invariance $s_1\, L_0 = {d}/{dt} \,(- W^{\prime} \,\psi)$ can be explained in 
the following manner in the language of the supervariables obtained after the application of SUSY 
restrictions [cf. (20)]. The translation of the super Lagrangian (31) along 
the direction of $\bar\theta$ is such that the result is a total derivative. In other words, the super 
Lagrangian (31) is a combination of composite supervariables, obtained after the application of SUSY 
restrictions, such that its translation along $\bar\theta$-direction
of the (1, 1)-dimensional super-submanifold produces a result which is nothing but the total time derivative.

 In exactly similar fashion, the starting Lagrangian (1) can {\it also} be expressed in terms of the chiral supervariables, 
obtained after SUSY restrictions  [cf. (28)], as 
\begin{eqnarray}
L_0 \; \Longrightarrow \; {\tilde L}^{(c)}_0 &=& \frac{1}{2}\,\dot {X^{(2)}}(t, \theta)\,\dot {X^{(2)}}(t, \theta)
 + i\, {\bar\Psi}^{(2)} (t, \theta)\,{\dot{\Psi}}^{(2)}(t, \theta)  \nonumber\\ 
&+& \frac{1}{2}\,{\tilde A}^{(2)}(t, \theta)\,{\tilde A}^{(2)}(t, \theta)
 + {\tilde W}^{\prime}(X^{(2)})\;{\tilde A}^{(2)}(t, \theta)
 \nonumber\\ &+& {\tilde W}^{\prime\prime}(X^{(2)}) \, {\bar\Psi}^{(2)} (t, \theta)\,{{\Psi}}^{(2)}(t, \theta),
\end{eqnarray}
where ${\tilde W}^{\prime}(X^{(2)})$ and ${\tilde W}^{\prime\prime}(X^{(2)})$ have the same expansions as quoted in (32)
with the replacements: $\bar\theta\,\rightarrow \,\theta$ and $\psi \rightarrow \bar\psi$. The invariance of the 
original Lagrangian (1) under $s_2$ can be captured in the following fashion:
\begin{eqnarray}
 \frac{\partial}{\partial \theta}\, {\tilde L}^{(c)}_0 = \frac{d}{dt} \,\Big[i\,\bar\psi\,(\dot x - i A - i\,W^{\prime}) \Big] 
 \Leftrightarrow  \; s_2\, L_0 = \frac{d}{dt} \,\Big[i\,\bar\psi\,(\dot x - i A - i\, W^{\prime})\Big].
\end{eqnarray}
Geometrically, the SUSY invariance of Lagrangian (1) is equivalent to the translation of the composite 
supervariables  (present in  ${\tilde L}^{(c)}_0$ [cf. (34)]) such that the outcome of the translation is a 
total derivative. Finally, we observe that the action integral can be expressed as:    
\begin{eqnarray}
S= \int dt\; L_0 \leftrightarrow   S=  \int dt \, {\tilde L}^{(ac)}_0 \leftrightarrow  
 S=  \int dt\, {\tilde L}^{(c)}_0, 
\end{eqnarray}
which is self-evident  from (31) and (34) because we observe that
${\tilde L}^{(ac)}_0 = L_0 + \bar\theta\, \frac{d}{dt} \,[-  W^{\prime}\,\psi ]$ and
${\tilde L}^{(c)}_0 = L_0 + \theta\, \frac{d}{dt} \,[i\,\bar\psi\,(\dot x - i A - i\, W^{\prime}) ]$. 
Thus, the inter-relationships, given in (36), are correct because the total derivative terms vanish due to Gauss's divergence theorem.

We can express the supercharge $Q$ in terms of the supervariables, 
obtained after the application of SUSY restrictions, in two different ways as: 
\begin{eqnarray}
Q &=&\frac{\partial}{\partial \bar\theta}\, \Big[- i\,\bar\Psi^{(1)}(t, \bar\theta)\,\Psi^{(1)}(t, \bar\theta)\Big] 
\nonumber\\ &\equiv & \int d\bar\theta\,
\Big[- i\,\bar\Psi^{(1)}(t, \bar\theta)\,\Psi^{(1)}(t, \bar\theta)\Big],\nonumber\\ 
Q &=&\frac{\partial}{\partial \bar\theta}\, \Big[\Big(\dot x(t) + i\,A(t)\Big)\, X^{(1)}(t, \bar\theta)\Big]  
\nonumber\\ &\equiv & \int d\bar\theta\,
\Big[\Big(\dot x(t) + i\,A(t)\Big)\, X^{(1)}(t, \bar\theta)\Big].   
\end{eqnarray}
In view of the mappings (20) and (29), the above charges can be {\it also} expressed as follows: 
\begin{eqnarray}
&&Q =s_1\, \Big[- i\,\bar\psi\,\psi\Big],  \quad\qquad Q =s_1\, \Big[(\dot x +i \,A)\,x\Big],  
\end{eqnarray}
which prove the nilpotency of the charge $Q$ in the language of the nilpotency of transformations (2) 
as well as in terms of the nilpotency ($\partial^2_{\bar\theta} = 0$) of the translational generator ($\partial_{\bar\theta}$).
This can be seen by $s_1\, Q = -i\,\{Q, \, Q\} =  0 $ and $\partial_{\bar\theta}\, Q = 0$
(by exploiting the relationships (38) and (37), respectively).

In exactly similar fashion, we can express the supercharge $\bar Q$ in terms of the supervariables (28), 
obtained after the application of SUSY invariant restrictions, in two different ways as illustrated below: 
\begin{eqnarray}
\bar Q &=& \frac{\partial}{\partial \theta}\, \Big[i\,\bar\Psi^{(2)}(t, \theta)\,\Psi^{(2)}(t, \theta)\Big] 
\nonumber\\ &\equiv & \int d\theta\,
\Big[i\,\bar\Psi^{(2)}(t, \theta)\,\Psi^{(2)}(t, \theta)\Big],  \nonumber\\
\bar Q &=& \frac{\partial}{\partial \theta}\, \Big[\Big(\dot x(t) - i\,A(t)\Big)\, X^{(2)}(t, \theta)\Big]  
\nonumber\\ &\equiv & \int d\theta\,
\Big[\Big(\dot x(t) - i\,A(t)\Big)\, X^{(2)}(t, \theta)\Big].
\end{eqnarray}
The above relationships can be re-expressed in terms of the ordinary 1D variables and transformations $s_2$ of (2) as follows
\begin{eqnarray}
&&\bar Q =s_2\, \Big[ i\,\bar\psi\,\psi\Big],   \qquad\qquad \bar Q =s_2\, \Big[(\dot x - i \,A)\,x\Big], 
\end{eqnarray}
which establish the nilpotency of $\bar Q$ in the ordinary space due to $s_2\, \bar Q = -i\, \{\bar Q, \, \bar Q\} =  0$.
In the superspace, we observe that $\partial_\theta\, \bar Q = 0$ due to the nilpotency ($\partial^2_\theta = 0$)
of translational generator $\partial_\theta$  along the Grassmannian direction 
$\theta$ of the (1, 1)-dimensional chiral supermanifold.  
Hence, we have proven the nilpotency property in a clear fashion.

\section{Off-Shell Nilpotent $\mathcal{N} = 2$ SUSY Transformations: Towards Cohomological Interpretation }

For the sake of completeness, we shall discuss here the mathematical implications  of the off-shell nilpotent ($s^2_1 = 0, \; s^2_2 = 0$)
$\mathcal{N} = 2$ SUSY transformations $s_1$ and $s_2$ in the language of de Rham cohomological operators of differential geometry
which have been discussed thoroughly  in [14].  We note that the Lagrangian (1) remains invariant under the following
{\it unique} discrete symmetry transformations [14]
\begin{eqnarray}
&& x \rightarrow -\, x, \qquad t \rightarrow - \,t, \qquad \psi \rightarrow 
+\,\bar\psi, \qquad \bar\psi \rightarrow -\,\psi, \nonumber\\
&& A \rightarrow -\, A, \qquad W' \rightarrow -\, W',\qquad W'' \rightarrow +\,\, W'',
\end{eqnarray}  
where there is an explicit presence of the time-reversal as well as reflection (i.e. parity) symmetries. 
Furthermore, the above transformations are {\it physically}
interesting because the superpotential $W(x)$ is {\it even} under parity [i.e. $W(-x) = W(x)$] 
which is required for the existence of square-integrable eigenfunctions. 
The above discrete  transformations are {\it unique} because we observe the validity of the
following  [14]
\begin{eqnarray}
&& s_2\, \Phi_1 = -\, *\, s_1\,*\,\Phi_1 , \quad s_1\, \Phi_1 =  *\, s_2\,*\,\Phi_1,  \quad
s_2\, \Phi_2 = + \,*\, s_1\,*\,\Phi_2, \nonumber\\ &&  s_1\, \Phi_2 = -\,*\, s_2\,*\Phi_2, \qquad
 \Phi_1 =x,\, A,\, W^{\prime},\, W^{\prime\prime}, \;\,\qquad   \Phi_2 = \psi,\,\bar\psi.
\end{eqnarray}  
The above relationships are the analogue of the relationship $\delta = \pm\,*\,d\,*$ of differential geometry 
where $d = dt\,\partial_t$ ($d^2 = 0$) is the exterior derivative and $\delta$ (with $\delta^2 = 0$)
is the co-exterior derivative. 
The ($\pm$) signs in (42) are dictated by two successive  operations  of  (41)
on the specific variable, namely;
\begin{eqnarray}
*\,(*\, \Phi_1) = + \, \Phi_1,\qquad\quad *\,(*\, \Phi_2) = - \, \Phi_2,
\end{eqnarray}
where ($*$) corresponds to the discrete symmetry transformations (41) and the generic 
variables $\Phi_1$ and $\Phi_2$ have been explained in (42).

Now we concentrate on the {\it physical} meaning of $d$ and $\delta$ in the language of the symmetry transformations.
It is straightforward to check that  the  continuous symmetry transformations ($s_1,\, s_2,\,s_\omega $) of Sec. 2 
satisfy the following algebra in their operator form, namely;
\begin{eqnarray}
&& s^2_1  = 0,\,\qquad s^2_2  = 0, \,\qquad \{s_1,\, s_2\} = s_\omega = (s_1 + s_2)^2, \nonumber\\
&& \big[s_\omega,\, s_1 \big] = 0,\qquad \quad [ s_\omega,\, s_2 ] = 0,\qquad \quad \{s_1,\,s_2\} \ne 0,
\end{eqnarray}
where the operator $s_\omega$ is defined modulo a factor of ($-\,2i$). Furthermore, we note that $s_\omega$
is like the Casimir operator because it commutes with $s_1$ and $s_2$. A close look at (44) exemplifies that this algebra  is reminiscent 
of the algebra satisfied by the de Rham cohomological  operators ($d,\, \delta,\, \Delta$) 
of differential geometry [17-21]. The latter algebra is [17-21]:
\begin{eqnarray}
&&d^2 = 0,\qquad \quad \delta^2 = 0, \qquad \quad \{d,\, \delta\} = \Delta = (d + \delta)^2,\nonumber\\
&&\big[\Delta,\, d \big] = 0, \;\;\qquad \quad \big[\Delta,\, \delta \big] = 0, \;\;\qquad \quad \{d,\, \delta\} = 0. 
\end{eqnarray}
Thus, we conclude that there exists one-to-one correspondence between the algebraic  structures of (44) and (45).
As a result, we have provided the physical meaning to the abstract mathematical properties associated
with the cohomological operators ($d,\, \delta,\, \Delta$) of differential geometry in the language of continuous symmetry transformations.

To summarize, we  add that the algebra (44) can be shown to be emulated by the 
conserved charges ($Q,\, \bar Q, Q_\omega$) (cf. Sec. 2) if we modify a bit the transformations 
(2) by an overall {\it constant} factor [14]. The other properties of ($d,\, \delta,\, \Delta$)
can be captured by the above charges where the eigenvalues and eigenfunctions 
(defined in the quantum Hilbert space) play important roles. Thus, ultimately, we observe that the general $\mathcal{N} = 2$ SUSY quantum mechanical model (with any arbitrary superpotential) 
provides the physical realizations of the cohomological operators. We wish to add that we have {\it not}
focused here on the formal mathematical quantities (see, e.g. [22-25])
 like the spin complex structure, $Z_2$-grading, Witten's parity 
operator, etc., in the discussion of our $\mathcal{N} = 2$ theory and its symmetries. We shall discuss about 
these formal aspects of the cohomological features in our future publication.

\section{Conclusions}

The main result of our present investigation is the derivation of the {\it full} set of off-shell nilpotent SUSY
symmetries ($s_1$ and $s_2$) [cf. (2)] for the general $\mathcal{N} = 2 $ SUSY QM model (with any arbitrary
superpotential $W(x)$) using the supervariable approach. We have defined the supervariables
[corresponding to the 1D ordinary variables of Lagrangian (1)] on the 
(1, 1)-dimensional (anti-)chiral super-submanifolds  of the general (1, 2)-dimensional supermanifold. 
It is the strength of the SUSYIRs on the (anti-)chiral supervariables that 
we have been able to derive the above SUSY transformations $s_1$ and $s_2$ accurately. Primarily, we have demanded that 
the SUSY invariant 1D quantities must remain independent of the ``soul" coordinates $\theta$ and $\bar\theta$  when the former are generalized onto the appropriate super-submanifolds. This requirement is {\it physically} cogent and logically appealing. It is pertinent to point out that, in the old literature (see, e.g. [19]),
the space coordinate $x(t)$ has been christened as the ``body" coordinate and the Grassmannian variables 
($\theta$ and $\bar\theta$) have been named as ``soul" coordinates. The former could be realized 
physically but the latter are totally mathematical and abstract in nature.
Thus, a SUSY invariant quantity must remain independent of the latter coordinates as they are only
mathematical artifacts.

Geometrically, we have shown that the translation of the supervariables, obtained after the application 
of SUSY invariant restrictions, along the Grassmannian  directions $\bar\theta$ and $\theta$ produces 
the SUSY transformations $s_1$ and $s_2$ (cf. Sec. 3 and 4). The nilpotency of $s_1$ and $s_2$ 
is deeply connected with two successive translations along the  Grassmannian directions 
$\bar\theta$ and $\theta$ which are generated by the nilpotent ($\partial^2_{\bar\theta} 
= \partial^2_{\theta} = 0$) translational generators $\partial_{\bar\theta}$ and $\partial_{\theta}$
on the (anti-)chiral (1, 1)-dimensional super-submanifolds. The symmetry invariance of the Lagrangian, 
under $s_1$ and $s_2$, is connected with the translation of some combination of  composite supervariables 
(obtained after SUSY invariant restrictions) along 
$\bar\theta$ and $\theta$-directions such that the outcome of these translations is a total 
time derivative in the ordinary one (0  + 1)-dimensional (1D) space.

One of the decisive features of our supervariable approach is the intelligent choice of (1, 1)-dimensional (anti-)chiral super-submanifolds
of the general (1, 2)-dimensional supermanifold on which the (anti-)chiral supervariables are defined. The latter are subjected to
the SUSY invariant restrictions which lead to the derivation  of  off-shell nilpotent $\mathcal{N} = 2$ SUSY transformations 
$s_1$ and $s_2$. The off-shell nilpotency property is encoded in the nilpotency ($\partial^2_\theta = 0,\,\partial^2_{\bar\theta} = 0$)
of the translational generators ($\partial_\theta$ or $\partial_{\bar\theta}$) along the $\theta$ or $\bar\theta$
direction of the (anti-)chiral (1, 1)-dimensional super-submanifolds (which accommodate
the existence of the above (anti-)chiral supervariables). This choice has been responsible in demonstrating that 
$s_1$ and $s_2$ {\it do} not respect the anticommutativity property (i.e. $\{s_1,\, s_2\} \ne 0$).

We hope to generalize  our analysis for the description of the extended 
SUSY quantum mechanical models (with $\mathcal{N} = 4, 6, 8....$). Furthermore, we can
implement the procedure of dimensional reduction to obtain the one (0 + 1)-dimensional 
$\mathcal{N} = 2$ and $\mathcal{N} = 1$ Yang-Mills models thereby establishing a connection with the SUSY gauge theories.
An excellent set of works [22-25] exist in this regard which we wish to apply in our future endeavors.
We are also trying to apply SUSY version of HC to obtain the proper
(anti-)BRST symmetries for the SUSY gauge theories  [26]. In this context,
we are sure that the idea of SUSY invariant restrictions would play very important role as they would 
provide additional useful restrictions. Thus, the methodology and ideas used in our present text would be useful in deriving
SUSY as well as (anti-)BRST symmetries for the SUSY gauge theories.
Currently, we are devoting time on it and our results would be 
reported elsewhere [27].

Before we close this section, we would like to briefly mention here the cohomological implications of
the ${\mathcal N} = 2$ SUSY quantum mechanical algebra. The presence of this algebra provides a $Z_2$-grading
of the quantum Hilbert space of states and it also generates transformations between even-odd parity states
(w.r.t. the Witten parity operator). This is why, the spacetime manifold 
turns out to be a {\it globally} graded manifold 
(but {\it not} a supermanifold). Thus, even though the cohomological operators $(d, \delta, \Delta)$
are identified with the  ${\mathcal N} = 2$ SUSY quantum mechanical charges $(Q, \bar  Q, H\equiv Q_{\omega})$, the
Hodge decomposition theorem can not be defined in the quantum Hilbert space of $Z_2$-graded quantum 
states. However, for the even {\it or} odd parity states of the total quantum Hilbert space, 
the (co-) cohomology w.r.t. the ${\mathcal N} = 2$ supercharges can be defined in the corresponding subspace. \\

\noindent
{\bf Acknowledgements:}
Two of us (SK and AS) would like to gratefully acknowledge the financial support
from UGC and CSIR, Government of India, New Delhi, under their SRF-schemes, respectively.
Enlightening comments by our esteemed Reviewer are thankfully acknowledged, too.\\

\noindent
\appendix
\section{On the Choice of (Anti-)chiral Supervariables}
\noindent
Here we explain the reasons  behind our intelligent  choice of the (anti-)chiral  supervariables in the context of derivation of the SUSY nilpotent transformations for the general $\mathcal{N} = 2$ SUSY quantum mechanical model of our present investigation. As pointed out (and emphasized in the main body of our text), the (anti-)BRST symmetry transformations for a gauge theory are nilpotent and absolutely anticommuting. Thus, corresponding 
to a given bosonic field $\phi (x)$ of this D-dimensional gauge theory, one has to generalize it onto a (D, 2)-dimensional supermanifold with the following general expansion  [4-7].
\[
\tilde \Phi (x, \theta, \bar\theta) =  \phi (x) + \theta \, \bar R(x)
 +\bar\theta\, R(x) + \theta\bar\theta\, S(x),  \hskip 6cm (A.1)\]
where $\tilde \Phi (x, \theta, \bar\theta)$ is the superfield defined on  the (D, 2)-dimensional
supermanifold and ($R(x), \bar R(x)$) are the fermionic secondary fields and $S(x)$ is a bosonic 
secondary field. As it turns out, the translational generators ($\partial_\theta, \partial_{\bar\theta}$) 
are found to correspond to the (anti-)BRST  symmetry transformations $s_{(a)b}$ which are 
nilpotent of order two due to ($\partial^2_\theta = \partial^2_{\bar\theta} = 0$) 
and they are absolutely anticommuting because it 
is straightforward to check that:
\[
\partial_{\bar\theta} \, \partial_{\theta}\,\Big(\tilde \Phi (x, \theta, \bar\theta)\Big) 
= i\, S(x)\quad \Longleftrightarrow \quad s_b \,s_{ab}\, \phi (x), \hskip 6cm (A.2)\]
\[
\partial_{\theta} \, \partial_{\bar\theta}\,\Big(\tilde \Phi (x, \theta, \bar\theta)\Big) 
= - i\, S(x) \quad \Longleftrightarrow \quad s_{ab} \,s_{b}\, \phi (x). \hskip 5.7cm (A.3)\]
Thus, we observe that $(s_b\, s_{ab} + s_{ab}\, s_b)\,\phi (x)  = 0$ due to 
the above relations (A.2) and (A.3) which are also implied by
$(\partial_{\bar\theta} \, \partial_{\theta}
+ \partial_{\theta} \, \partial_{\bar\theta}) = 0$.
In our present investigation (connected with the SUSY QM theory), 
we are compelled  to avoid relations of the type (A.2) and (A.3) so that our nilpotent SUSY 
symmetries {\it could not} become absolutely 
anticommuting in nature.

We wrap up this Appendix A  with the remarks that our SUSY nilpotent symmetries are geometrically identified 
with the translational generators ($\partial_\theta, \partial_{\bar\theta}$) along the Grassmannian  
directions of the (anti-)chiral super-submanifolds  which encapsulate
only the {\it nilpotency} of the symmetry transformations.
The {\it above choice} also makes it clear that one can derive both 
the $\mathcal{N}  = 2$ SUSY symmetries  
{\it independently}. As far as the computation  of the anticommutativity property 
is concerned, one has to compute it  and check its nature {\it separately}
 after  derivation of the $\mathcal{N}  = 2$ SUSY symmetries by our proposed method  
of supervariable approach.
 \\

\end{document}